\documentclass[pra,preprint,showpacs,amsmath,amssymb,superscriptaddress]{revtex4}
\usepackage{graphicx}
\usepackage{natbib}
\usepackage{latexsym}

\begin{document}

\title{Controlling the sense of molecular rotation: classical vs quantum analysis}

\author{Yuri Khodorkovsky}
\affiliation{Department of Chemical Physics, The Weizmann Institute of Science, Rehovot 76100, Israel}

\author{Kenta Kitano}
\altaffiliation[Present address: ]{Institute for Solid State Physics, The University of Tokyo, 5-1-5 Kashiwanoha, Chiba 277-8581, Japan}
\affiliation{Institute for Molecular Science and SOKENDAI (The Graduate University for Advanced Studies) Myodaiji, Okazaki 444-8585, Japan}

\author{Hirokazu Hasegawa}
\altaffiliation[Present address: ]{Dept.\ of Basic Science, The University of Tokyo, 3-8-1 Komaba, Tokyo 153-8902, Japan}
\affiliation{Institute for Molecular Science and SOKENDAI (The Graduate University for Advanced Studies) Myodaiji, Okazaki 444-8585, Japan}

\author{Yasuhiro Ohshima}
\affiliation{Institute for Molecular Science and SOKENDAI (The Graduate University for Advanced Studies) Myodaiji, Okazaki 444-8585, Japan}

\author{Ilya Sh.\ Averbukh}
\affiliation{Department of Chemical Physics, The Weizmann Institute of Science, Rehovot 76100, Israel}

\begin{abstract}
Recently, it was predicted theoretically and verified experimentally  that a pair of  delayed and cross-polarized short laser pulses can create molecular ensembles with a well defined sense of rotation (clockwise or counterclockwise). Here we provide a comparative study of the classical and quantum aspects of the underlying mechanism for linear molecules and for symmetric tops, like benzene molecules, that were used for the first experimental demonstration of the effect. Very good quantitative agreement is found between the classical description of the process and the rigorous quantum mechanical analysis at the relevant experimental conditions. Both approaches predict the same optimal values for the delay between pulses and the angle between them, and deliver the same magnitude of the induced oriented angular momentum of the molecular ensemble. As expected, quantum and classical analysis substantially deviate when the delay between pulses is comparable with the period of quantum rotational revivals. However,   time-averaged  characteristics of the excited  molecular ensemble  are equally good described by the these two approaches. This is illustrated by calculating the anisotropic time-averaged angular distribution of the double-pulse excited molecules, which reflects persistent confinement of the molecular axes to the rotation plane defined by two polarization vectors of the pulses.

\end{abstract}
\maketitle

\section{Introduction}

Laser control of molecular rotation, alignment and orientation has received significant attention in recent years (for a review, see e.\ g.\ \cite{Stapelfeldt,Tamar}). Interest in the field has increased, mainly due to the improved capabilities to manipulate characteristics of the laser pulses (such as time duration and temporal shape), which in turn leads to potential applications offered by controlling the angular distribution of molecules. Since the typical rotational motion is 'slow' (${\sim}10 \,\mathrm{ps}$) with respect to the typical short pulse (${\sim}50 \,\mathrm{fs}$), effective rotational control and manipulation are in reach.
During the last decade, temporal rotational dynamics of pulse-excited molecules was studied \cite{Ortigoso,Rosca-Pruna}, and multiple pulse sequences giving rise to the enhanced alignment were suggested \cite{Averbukh,Leibscher,Renard} and realized \cite{Bisgaard1,Lee,Bisgaard2,Pinkham}. Further manipulations such as optical molecular centrifuge and alignment-dependent strong field ionization of molecules were demonstrated \cite{Karczmarek,Litvinyuk}. Selective rotational excitation in bimolecular mixtures was suggested and demonstrated in the mixtures of  molecular isotopes \cite{Fleischer1} and molecular spin isomers \cite{Faucher,Fleischer2}.

\begin{figure}
\centering
\includegraphics[width=1\textwidth]{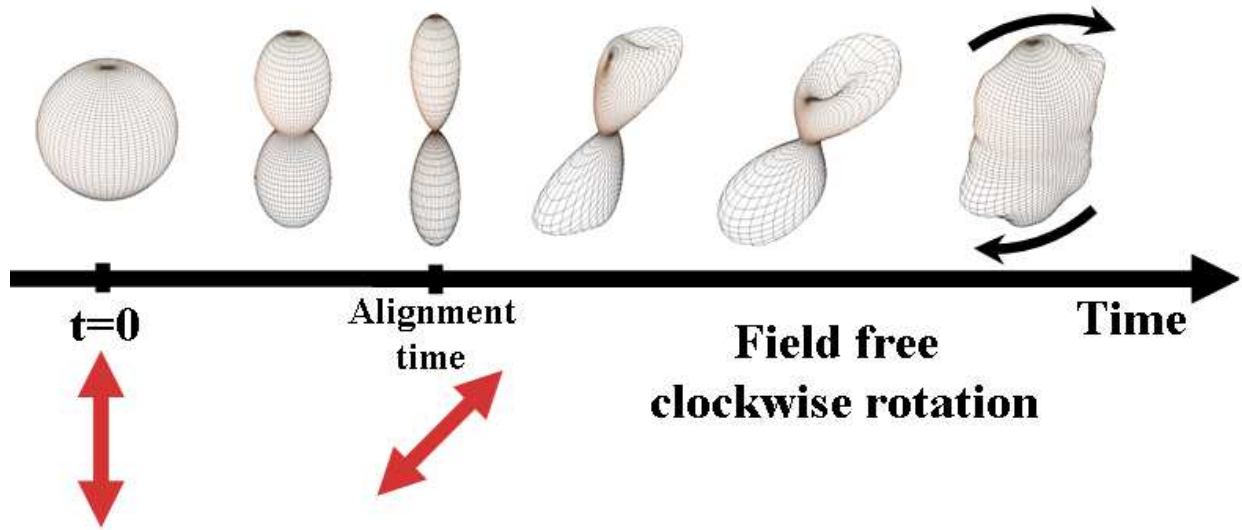}
\caption{(Color online) ``Molecular propeller'' (figure taken from reference \cite{njp}).}
\label{propeller}
\end{figure}

Recently, several groups suggested a method for exciting field-free unidirectional molecular rotation, in which the angular momentum along the laser beam propagation direction is provided by two properly delayed ultrashort laser pulses that are linearly polarized at 45 degrees to each other \cite{ultrafast,cleo,njp,York,Kitano}. The first experimental demonstration of the effect was done in \cite{Kitano} for benzene molecules. The mechanism behind this double-pulse scheme is rather clear, and is most easily explained in the case of linear molecules \cite{ultrafast,cleo,njp}. The first ultrashort laser pulse, linearly polarized along the $z$-axis, impulsively induces coherent molecular rotation that continues after the end of the pulse.  The molecules rotate under field-free conditions until they reach an aligned state, in which the molecular axis with the highest polarizability is  confined in a narrow cone around the polarization direction of the first pulse. The second short laser pulse is applied at the moment of the best alignment, and at an angle with respect to the first pulse. As a result, the aligned molecular ensemble experiences a torque causing molecular rotation in the plane defined by the two polarization vectors. The torque acting on a linear molecule is maximal when the laser pulse is polarized at 45 degrees with respect to the molecular axis of the highest polarizability. This defines the optimal angle between the  laser pulses. The direction of the excited rotation (clockwise or counter-clockwise) is determined by the sign of the relative angle ($\pm45$ degrees) between the first and the second pulse in the polarization plane. This double pulse scheme (see Fig.\ \ref{propeller}) was termed ``molecular propeller'' \cite{ultrafast,cleo,njp}, as it resembles the action (side kick) needed to ignite a rotation of a plane propeller.
The heuristic arguments presented above are, in fact, essentially classical, although it is clear that a detailed quantum analysis may be needed to consider the long-time evolution of the excited molecules, or the operation of this scheme at relatively weak pulses when only low-lying rotational states are involved. That is why all the previous treatments of the problem \cite{ultrafast,cleo,njp,York,Kitano} presented full-scale quantum-mechanical studies, combining both analytical and numerical methods.

In the present paper, we provide a detailed comparison between the classical and the quantum-mechanical analysis of the double pulse scheme for exciting unidirectional molecular rotation, both for linear molecules and for benzene (that was the object of the experiment in \cite{Kitano}). We will demonstrate the predictive power of the classical treatment that correctly prescribes the optimal delay between the pulses, and the angle between them, and will find the advantages and limitations of such an approach in quantitative description of the effect at experimental conditions. We believe that by comparing the complementary classical and fully quantum mechanical views on the same phenomenon, the reader will get a comprehensive idea about the underlying physics.

Recent paper \cite{njp} studied quantum-mechanically  the ``molecular propeller'' scheme in the case of linear molecules. Dependence of the induced angular momentum  on the pulses' intensity, and the delay between them, was analyzed. Moreover, the  angular distribution of linear molecules subject to unidirectional rotation was considered, and it was shown to be confined in the plane defined by the two polarization vectors of the pulses. This anisotropic angular distribution was characterized by the observable $\langle\cos^2{\varphi}\rangle$, which was referred to as the azimuthal factor.

In the present paper, we describe the same scheme using a classical description of the double-pulse molecular excitation, followed by a Monte Carlo averaging over the thermal ensemble. We compare the time dependence of the azimuthal factor according to the classical analysis, to the quantum mechanical one, and show that both results coincide for initial times. In addition, we introduce a long-time-averaged angular distribution function, which conveniently depicts the molecular orientation following the excitation. This distribution function seems to behave similarly in the classical and quantum cases. Finally, we analyze classically the excitation of the unidirectional rotation for more complicated oblate symmetric top molecules, like benzene,  and compare this analysis with the full quantum-mechanical treatment of the problem.

The outline of the paper is as following. In Sec.\ II we study classically the rotational motion of a  linear molecule kicked by a pair of short nonresonant laser pulses that are linearly polarized in  arbitrary directions. In Sec.\ III,  we describe the response of a thermal ensemble of linear molecules to such an excitation by using the Monte Carlo method. Angular distribution function and its time-averaged behavior are studied in Sec.\ IV, and the results are compared to the full quantum mechanical treatment. In Sec.\ V, we analyze (both classically and quantum mechanically) the problem of inducing unidirectional rotation of benzene molecules.  Sec.\ VI summarizes and concludes our study.

\section{Classical dynamics of a kicked linear molecule}

Here we consider classical dynamics of a model diatomic (or, more generally, linear) polarizable molecule that interacts with an ultrashort laser pulse. The molecule is treated as a linear rigid rotor.
We assume that the central frequency of the pulse is far detuned from any molecular transition. The potential energy  of the pulse interaction with the induced molecular polarization is given by \cite{Boyd}:
\begin{equation}
\label{V}
V(\theta,\varphi,t)=-\frac{1}{4}\mathcal{E}^2(t)\left(\Delta\alpha\cos^2{\beta}+\alpha_{\perp}\right)~,
\end{equation}
where $\Delta\alpha=\alpha_{\parallel}-\alpha_{\perp}$ is the difference between the polarizability along the molecular axis and the one perpendicular to it, $\mathcal{E}(t)$ is the envelope of the electric field of the {\it linearly polarized} laser pulse, and $\beta=\beta(\theta,\varphi)$ is the angle between the molecular axis and the direction of polarization of the pulse. Here $\theta,\varphi$ are the polar and the azimuthal angles characterizing the direction of the molecular axis, respectively.  We assume the pulse duration is very short compared to the rotational period, so that the pulse can be described in the impulsive ($\delta$-kick) approximation. We define the dimensionless interaction strength $P$, which characterizes the pulse, as
\begin{equation}
\label{P}
P=\frac{\Delta\alpha}{4\hbar}\int_{-\infty}^{\infty}\mathcal{E}^2(t)dt~.
\end{equation}

Let us remind again the pulse sequence producing unidirectionally rotating molecules \cite{ultrafast,cleo,njp}: (1) at time $t{=}0$  molecules are kicked by a pulse linearly polarized along the $z$-axis; (2) at the time of the maximal molecular alignment ($t{=}t_{\text{al}}$, which is the moment when the alignment factor $\langle\cos^2{\theta}\rangle$ reaches the maximum value), a second pulse is applied, polarized at 45 degrees to the $z$-axis (we  choose the polarization vector to be in the $x$-$z$ plane). We represent the molecule by a unit vector pointing from the center of mass towards one of the atoms (the center of mass is defined as the coordinates' origin). The tip of the vector defines a point on the surface of the unit sphere. The orientation of the unit vector is given by
\begin{equation}
\label{r}
\mathbf{r}=\left(x,y,z\right)=\left(\sin{\theta}\cos{\varphi},\sin{\theta}\sin{\varphi},\cos{\theta}\right)~.
\end{equation}
The rotational velocity of the unit vector $\mathbf{r}$ can be described by two orthogonal angular velocities $v_{\theta}=\dot{\theta}$ and $v_{\varphi}=\dot{\varphi}\sin{\theta}$ (the dot denotes a derivative with respect to time), or, in vector notation,
\begin{equation}
\label{v_spher2cart}
\mathbf{v}=v_{\theta}\vec{e}_{\theta}+v_{\varphi}\vec{e}_{\varphi}~,
\end{equation}
where the spherical unit vectors are
\begin{eqnarray}
\vec{e}_{\theta}&=&\left(\cos{\theta}\cos{\varphi},\cos{\theta}\sin{\varphi},-\sin{\theta}\right) ~~;\nonumber \\
\label{unit_vec}
\vec{e}_{\varphi}&=&\left(-\sin{\varphi},\cos{\varphi},0\right)~.
\end{eqnarray}
This rotational velocity is referred to as simply ``velocity'' in the following.
The rotational kinetic energy in this notation is given by
\begin{equation}
\label{kinetic_en}
T=\frac{1}{2}I\left(v_{\theta}^2+v_{\varphi}^2\right)~,
\end{equation}
where $I$ is the moment of inertia of the molecule.

As the first step, we find the change in the molecular velocity due to interaction with a single short laser pulse.
For a  $z$-polarized pulse, angle $\beta$ is equal to the polar angle $\theta$, and we can use the potential energy from Eq.\ (\ref{V})
to find the torque $\mathbf{\Gamma}$ applied to the molecule in spherical coordinates:
\begin{equation}
\label{force}
\Gamma_{\theta}=-\frac{\partial V}{\partial\theta}=-\frac{\Delta\alpha}{4}\mathcal{E}^2(t)\sin{2\theta} ~~;~~ \Gamma_{\varphi}=-\frac{\partial V}{\partial\varphi}=0
\end{equation}
Newton's second law gives $\Gamma_{\theta}=I\dot{v}_{\theta}=I\ddot{\theta}$ and $\Gamma_{\varphi}=I\dot{v}_{\varphi}=I\frac{d}{dt}(\dot{\varphi}\sin{\theta})$. We assume that the molecule does not change its orientation during the short pulse (impulsive approximation). By integrating the Newton's equations of motion over the short duration of the pulse, we obtain the change of the velocity due to the laser pulse action:
\begin{equation}
\label{delta_v}
\Delta v_{\theta}=-\frac{\hbar}{I}P\sin{2\theta_0} ~~;~~ \Delta v_{\varphi}=0
\end{equation}
Integrating again, and assuming that the molecule was initially at rest, we get
\begin{equation}
\label{theta_phi}
\theta(t)=\theta_0-\frac{\hbar}{I}P t\sin{2\theta_0} ~~;~~ \varphi=\varphi_0~.
\end{equation}
Here $\theta_0$ and $\varphi_0$ describe the initial orientation of the molecule. Every molecule starts to rotate with a constant velocity (depending on $\theta_0$ and $\varphi_0$ ), and $\mathbf{r}$ is tracing  a circle passing through the north and the south poles of the unit sphere.

\begin{figure}
\centering
\includegraphics[width=0.5\textwidth]{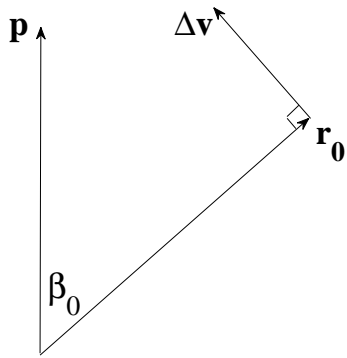}
\caption{A linear molecule, represented by $\mathbf{r_0}$, is kicked by a laser pulse linearly polarized along an arbitrary direction $\mathbf{p}$. As a result, it receives a velocity $\boldsymbol{\Delta}\mathbf{v}$.}
\label{kick_picture}
\end{figure}

As the next step, we consider the action of a pulse linearly polarized along some arbitrary unit vector $\mathbf{p}$, and determine the vector of the resulting velocity change $\boldsymbol{\Delta}\mathbf{v}$ for a molecule oriented along some direction $\mathbf{r_0}$. The norm $|\boldsymbol{\Delta}\mathbf{v}|$ can be found similarly to the previous case of a $z$-polarized pulse, and it is equal to $\frac{\hbar}{I}|P\sin{2\beta_0}|$, where $\beta_0$ is the angle between the polarization direction of the pulse $\mathbf{p}$ and the initial orientation direction of the molecule $\mathbf{r_0}$).  The direction of $\boldsymbol{\Delta}\mathbf{v}$ can be defined from the vector diagram in Fig.\ \ref{kick_picture}. Notice that $\boldsymbol{\Delta}\mathbf{v}$ is always perpendicular to $\mathbf{r_0}$. Also, $\boldsymbol{\Delta}\mathbf{v}$ is directed parallel or antiparallel to the vector component of $\mathbf{p}$ perpendicular to $\mathbf{r_0}$, which is equal to $\mathbf{p}-\cos{\beta_0}\mathbf{r_0}$. As a result, we arrive at:
\begin{equation}
\label{delta_v_vec}
\boldsymbol{\Delta}\mathbf{v}=\frac{2\hbar P}{I}\cos{\beta_0}\left(\mathbf{p}-\cos{\beta_0}\mathbf{r_0}\right)~.
\end{equation}
We can easily check that (\ref{delta_v_vec}) gives the correct result for $z$-polarized pulse, for which $\mathbf{p}=(0,0,1)$, by transforming (\ref{delta_v}) to cartesian coordinates using Eq.\ (\ref{v_spher2cart}).

Finally, we provide a formula for the time dependence of the position $\mathbf{r}$ and  velocity $\mathbf{v}$ of the point representing molecular orientation on the unit sphere during the field-free rotation, given the initial orientation $\mathbf{r_0}$ and initial velocity $\mathbf{v_0}$. As $\mathbf{v}$ is always perpendicular to $\mathbf{r}$,  the tip of the $\mathbf{r}$ vector  traces a  circle on the sphere. The  orientation of the molecule at time $t$ is
\begin{equation}
\label{r_of_t}
\mathbf{r}(t)=\mathbf{r_0}\cos{(v_0t)}+\frac{\mathbf{v_0}}{v_0}\sin{(v_0t)}~,
\end{equation}
where $v_0=|\mathbf{v_0}|$. Taking the derivative, we find the molecular velocity:
\begin{equation}
\label{v_of_t}
\mathbf{v}(t)=-v_0\mathbf{r_0}\sin{(v_0t)}+\mathbf{v_0}\cos{(v_0t)}~.
\end{equation}

\section{Ensemble averaging and Monte Carlo simulation}

We are interested in the response of a thermal molecular ensemble to the double-pulse excitation, and in the calculation of ensemble averaged values of the alignment factor and other observable quantities. One way to calculate the time dependence of an observable averaged over the ensemble is to find its value at time $t$ using (\ref{theta_phi}), or (\ref{r_of_t}), and to integrate it over the initial angular distribution function \cite{Leibscher}. For example, at {\it zero temperature}, the molecules are initially at rest and isotropically distributed in all directions. Therefore, for any observable $f(\theta,\varphi)$, we have to substitute the time-dependent expressions (\ref{theta_phi}), (\ref{r_of_t}) for $\theta$ and $\varphi$  to obtain $f(\theta_0,\varphi_0,t)$, and to integrate over the isotropic distribution of $\theta_0$ and $\varphi_0$:
\begin{equation}
\label{avg_over_ens_0temp}
\langle f(\tau)\rangle=\frac{1}{4\pi}\int\limits_0^{2\pi}d\varphi_0\,\int\limits_0^{\pi} d\theta_0\sin{\theta_0}\,f(\theta_0,\varphi_0,\tau)~.
\end{equation}
Here the brackets denote averaging over the ensemble. For a {\it nonzero temperature}, an averaging over the initial velocity distribution should be performed as well. In a thermal ensemble at temperature $T$, the velocities $v_{\theta,0}$ and $v_{\varphi,0}$ are distributed according to the Boltzmann distribution:
\begin{eqnarray}
&\mathrm{Prob}(v'_{\theta,0},v'_{\varphi,0}) dv'_{\theta,0} dv'_{\varphi,0} \nonumber \\
&=\frac{1}{\sqrt{2\pi\sigma_{\text{th}}^2}}\exp{\left[-\frac{\left(v'_{\theta,0}\right)^2}{2\sigma_{\text{th}}^2}\right]} dv'_{\theta,0}  \nonumber \\
\label{boltzmann}
&\times  \frac{1}{\sqrt{2\pi\sigma_{\text{th}}^2}} \exp{\left[-\frac{\left(v'_{\varphi,0}\right)^2}{2\sigma_{\text{th}}^2}\right]} dv'_{\varphi,0}~,
\end{eqnarray}
where the dimensionless width of the thermal distribution $\sigma_{\text{th}}$ is given by $\sigma_{\text{th}}^2=Ik_BT/\hbar^2=(2hBc/k_BT)^{-1}$. Here $B$ is the rotational constant of the molecule, and  $v'$ is the dimensionless velocity $Iv/\hbar$. Instead of (\ref{avg_over_ens_0temp}), we have
\begin{eqnarray}
\label{avg_over_ens_non0temp}
\langle f(\tau)\rangle&=&\frac{1}{4\pi}\int\limits_0^{2\pi}d\varphi_0\,\int\limits_0^{\pi} d\theta_0\sin{\theta_0}\, \int\limits_{-\infty}^{\infty} dv'_{\theta,0}  \\
~&~&\int\limits_{-\infty}^{\infty}\,  dv'_{\varphi,0}\, \mathrm{Prob}(v'_{\theta,0},v'_{\varphi,0}) f(\theta_0,\varphi_0,v'_{\theta,0},v'_{\varphi,0}, t)~. \nonumber
\end{eqnarray}
These integrals have fast oscillating integrands and are difficult (although possible) to calculate numerically.

A somewhat simpler way of finding the time dependence of observables averaged over the molecular ensemble is by means of a Monte Carlo simulation. We start with molecules isotropically oriented in space, with rotational velocities distributed according to Eq.\ (\ref{boltzmann}). These initial values are obtained using a numerical random number generator. In order to randomly distribute points on the surface of the sphere, one uses the probability function
\begin{eqnarray}
\mathrm{Prob}(\theta,\varphi)d\theta d\varphi&=&\mathrm{Prob}(\theta)d\theta \times \mathrm{Prob}(\varphi)d\varphi \nonumber \\
\label{rand_r_sphere}
~&=&\frac{1}{2}\sin{\theta}\,d\theta \times \frac{1}{2\pi}d\varphi~.
\end{eqnarray}
The realization of $\mathrm{Prob}(\varphi)$ is straightforward if one has a {\it uniform} random number generator. If $w_{\text{un}}$ denotes a random number from a uniform distribution in the interval $[0,1]$, then $\varphi_0^{\text{rand}}=2\pi w_{\text{un}}$. On the other hand, in order to realize $\mathrm{Prob}(\theta)$ one should use the transformation method \cite[section 7.2]{numerical_recipes}. Starting with the equality of probabilities $dw=\frac{1}{2}|\sin{\theta}d\theta|$, we obtain $\theta_0^{\text{rand}}=2\arcsin{\sqrt{w_{\text{un}}}}$.

\begin{figure}
\centering
\includegraphics[width=1\textwidth]{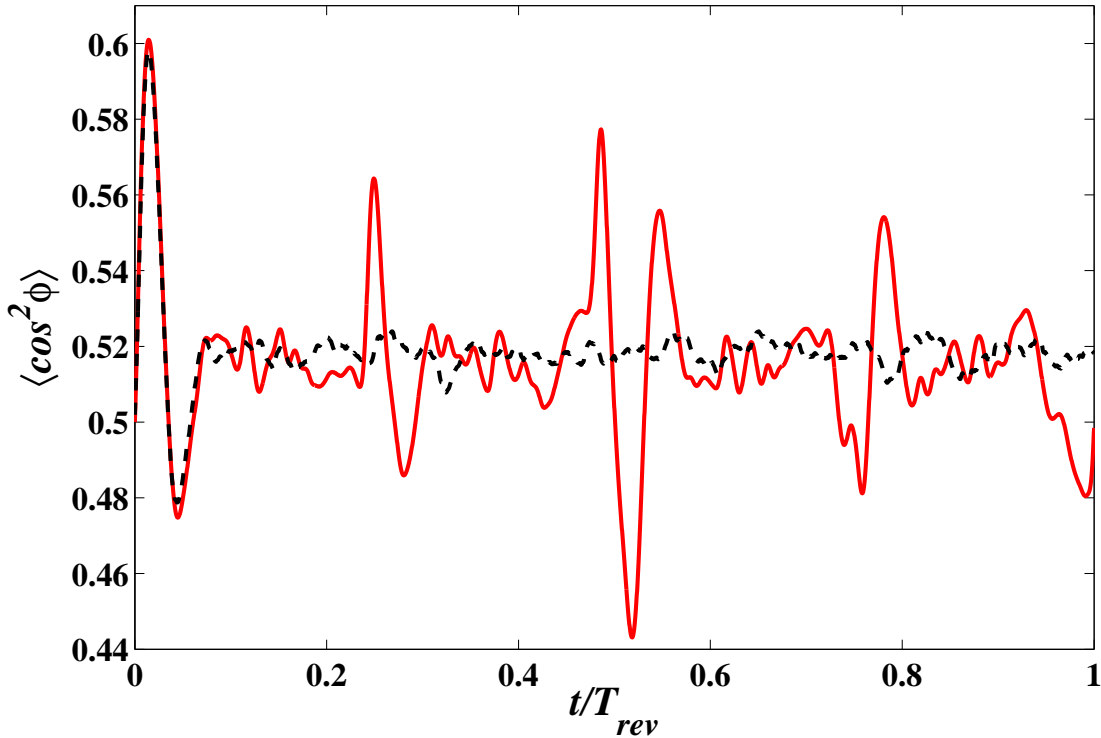}
\caption{(Color online) The ensemble-averaged azimuthal factor $\langle\cos^2{\varphi}\rangle$ as a function of time after the second pulse exciting unidirectional molecular rotation. It is calculated by the classical Monte Carlo method with $10^4$ molecules (dashed black), and by the exact quantum (finite-difference time-domain) simulation (solid red). The polarization vectors are $\mathbf{p_1}=(0,0,1)$ and $\mathbf{p_2}=(1,0,1)/\sqrt{2}$, and $P=5$ for both pulses. The calculation was done for nitrogen molecules at the rotational temperature of $50\,\mathrm{K}$. The time is measured in units of the quantum revival time $T_{\text{rev}}=2\pi I/\hbar$.}
\label{cos2phi}
\end{figure}

In order to realize the velocity distribution (\ref{boltzmann}), one uses the {\it normal} random number generator. Using the transformation method again, we obtain $v_{\theta,0}^{\text{rand}}=\sigma_{\text{th}}w_{\text{norm}}$ and $v_{\varphi,0}^{\text{rand}}=\sigma_{\text{th}}w_{\text{norm}}$, where $w_{\text{norm}}$ denotes a random number from a normal distribution with zero mean and standard deviation of unity.

Now we have all the ingredients needed to simulate the scheme for exciting unidirectionally rotating molecules, as explained at the beginning of Sec.\ II. Randomly oriented molecules are ``kicked'' by a pulse polarized along the $z$-axis, and get velocity additions according to Eq.\ (\ref{delta_v_vec}), which should be added to their initial thermal velocities. The molecules then evolve freely according to Eqs.\ (\ref{r_of_t}) and (\ref{v_of_t}). At each time step, we calculate the ensemble averaged alignment factor $\langle\cos^2{\theta}\rangle$. At the time of the maximal alignment, the 45-degrees rotated pulse is applied, velocities are changed again according to Eq.\ (\ref{delta_v_vec}), and another free propagation follows.

As the first application of this approach, we investigate the azimuthal factor $\langle\cos^2{\varphi}\rangle$ that characterizes confinement of the molecular angular distribution to the $x$-$z$ plane. This factor takes the unity value when all the molecules are confined to  this plane, and it is equal to $0.5$ for the isotropic ensemble.
In Figure \ref{cos2phi}, we plot the classically calculated azimuthal factor $\langle\cos^2{\varphi}\rangle$ as a function of time following the second pulse, along with the exact quantum result (for details of the quantum calculation see \cite{njp}). The calculation here (and also at Figs.\ \ref{z_kick_dist} and \ref{z_then45_dist} below) is done for nitrogen molecules with the rotational constant of $B=h/(8\pi^2Ic)=2.00\,\mathrm{cm}^{-1}$. The rotational temperature is $50\,\text{K}$, or $\sigma_{\text{th}}=2.94$.

It is clearly seen that the classical calculation (Monte Carlo realization using $10^4$ molecules) perfectly reproduces the first oscillation of the azimuthal factor. Obviously, later fractional and full revivals are quantum features, which are not reproduced by the classical treatment. It is also evident that the time averaged azimuthal factor (that  is around 0.52 here) coincides for the classical and the quantum calculation. This value is higher than the isotropic value of 0.5, hence the molecular distribution is squeezed towards the plane defined by the polarization directions of the first and the second pulses. We see that a relatively simple classical model reproduces the important features of this problem.

\section{Angular distribution function}

In this Section we  estimate classically the angular distribution function of an ensemble of linear molecules excited by two pulses, and compare it with the results of full quantum-mechanical simulation \cite{njp}. To achieve this goal, we use the kernel estimator method \cite{density} to reconstruct the distribution function from a set of $N>>1$ points on the unit sphere, which are participating in the Monte Carlo simulation. According to this method, each one of the $N$ points is surrounded by a narrow symmetric Gaussian function on the surface of the sphere. The sum of all Gaussians  represents the angular distribution function. Its time dependence is determined numerically by propagating independently $N$ rotating molecules, as  explained in the previous sections.

A simple way to build a symmetric Gaussian function around a point $\mathbf{r_i}(t)$ on a sphere  ($i$ denotes the molecule number in the Monte Carlo realization) is by referring to the angle $\alpha$ between the unit vector to this point, and the unit vector to a general point $\mathbf{r}$ in its neighborhood:  $\cos{\alpha}=\mathbf{r}\cdot \mathbf{r_i}(t)$. The corresponding (unnormalized) distribution function is $\exp{(-\alpha^2/2\sigma^2)}$, where $\sigma$ is the distribution width (a free parameter). Because only small width values are used ($\sigma\ll 1$), the quantity $\alpha^2/2$ can be replaced by the expression $1-\cos{\alpha}$, which is equivalent to it for $\alpha\ll 1$, while for $\alpha > 1$ the Gaussian distribution function is negligible any way.  Normalizing the two dimensional surface Gaussian function, we finally obtain the following estimate for the time dependent distribution function:
\begin{equation}
\label{TD_distribution}
\rho\left(\theta,\varphi,t\right)=\frac1N\sum_{i=1}^{N}\frac{1}{2\pi\sigma^2}\exp{\left(-\frac{1-\mathbf{r}\cdot\mathbf{r_i}(t)}{\sigma^2}\right)}~.
\end{equation}

An important object showing the symmetry and  anisotropy of our system is the time-averaged angular distribution function. A straightforward, but somewhat cumbersome way to generate it is by  numerical averaging the distribution function (\ref{TD_distribution}) over a long time period. A much simpler approach to time averaging is described below. After the end of the  pulses, each molecule moves with a constant speed on some circle on the surface of the unit sphere. Within a sufficiently long time period, each molecule spends the same amount of time at each length element of the circle (apart from very rare cases in which a molecule stays strictly at rest). Thus, the long-time-averaged angular distribution function for each molecule is a ``belt'' (a  circle with a width) on the unit sphere. Its orientation depends on the initial position $\mathbf{r_0}$ and velocity ${\mathbf{v_0}}$ on the unit sphere, and its width is a free parameter $\sigma_{\text{belt}}$, which, for consistency, is taken equal to the parameter $\sigma$ that was introduced before. The total time-averaged distribution function is obtained, as before, by adding  the ``belts'' for all the molecules participating in the Monte Carlo simulation.

The orientation of the ``belt'' for the $i$-th molecule  is given by the angular momentum unit vector $\vec{e}_{\mathbf{L}}=\mathbf{r_{0,i}}\times \mathbf{v_{0,i}}/v_{0,i}$, which is perpendicular to the plane defining the circle of rotation. Thus, the Gaussian ``belt'' for a single molecule is located close to the points $\mathbf{r}$ on the unit sphere, for which the dot product $\vec{e}_{\mathbf{L}}\cdot \mathbf{r}$ is nearly zero. Inserting the correct normalization factors, we finally obtain for the time-averaged angular distribution function, $\rho_{ta}$:
\begin{eqnarray}
\label{time_avg_distribution}
\rho_{\text{ta}}\left(\theta,\varphi\right)&=&\overline{\rho\left(\theta,\varphi,t \right)}\\
~&=&\frac1N\sum_{i=1}^{N}\frac{1}{2\pi\sqrt{2\pi\sigma_{\text{belt}}^2}}\exp{\left(-\frac{\left(\vec{e}_{\mathbf{L}} \cdot \mathbf{r}\right)^2}{2\sigma_{\text{belt}}^2}\right)}~,\nonumber
\end{eqnarray}
where the overline denotes a long time average.

\begin{figure}
\centering
\includegraphics[width=1\textwidth]{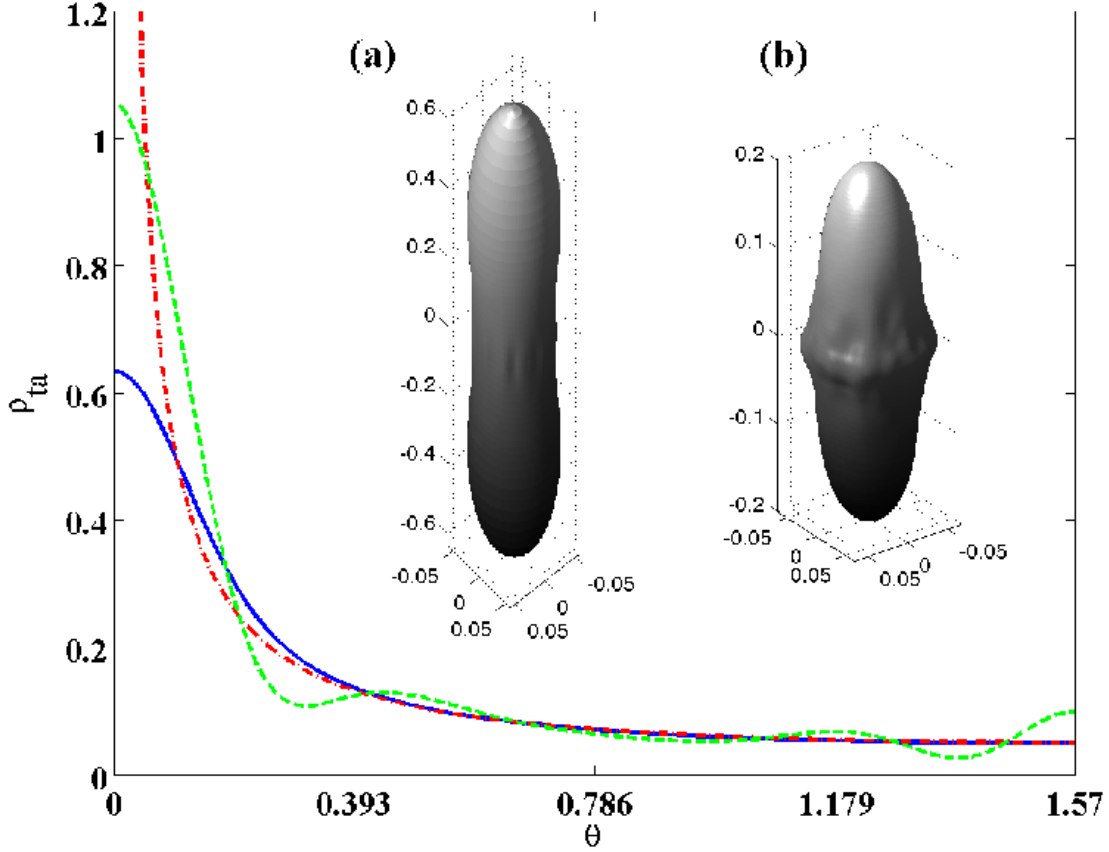}
\caption{(Color online) Time-averaged angular distribution function for an ensemble of nitrogen molecules kicked by a laser pulse with $P=10$. The curve plots and inset (a) are for molecules initially at zero rotational temperature. The dashed green curve presents the exact quantum result, and the dash-dotted red line displays the analytical classical distribution according to Eq.\ (\ref{dist_analytic}). The blue solid curve and inset (a) are calculated by the Monte Carlo method  according to Eq.\ (\ref{time_avg_distribution}), with $\sigma_{\text{belt}}=0.1$ and $N=10^4$. Inset (b) shows Monte Carlo results for the initial rotational temperature of $50\,\text{K}$. }
\label{z_kick_dist}
\end{figure}

Before considering the double pulse scheme, we first analyze the time-averaged angular distribution function for molecules subject to a single pulse polarized along the $z$-axis.
The result for zero temperature (all molecules are initially at rest) is shown in Fig.\ \ref{z_kick_dist} (a) and duplicated at the plot by the solid  line. In Fig.\ \ref{z_kick_dist} (b), the distribution function is plotted for the same system, but at finite temperature. As expected,  the distribution is ``cigar''-shaped, and azimuthally symmetric.
At zero temperature, an analytical result can be obtained for the time-averaged distribution function in the classical limit. In this case, all the trajectories of the kicked molecules are circles passing through the poles of the unit sphere. These circles are uniformly distributed in the azimuthal direction, and the distribution function is $\varphi$-independent. Therefore, at any given value of $\theta$ we find the same total number of molecules. As a result, the time-averaged distribution function multiplied by the geometrical factor $\sin{\theta}$ should be a constant. It is seen from here, that the distribution function averaged over long time should be proportional to $1/\sin{\theta}$, or, after normalization:
\begin{equation}
\label{dist_analytic}
\rho_{\text{ta}}\left(\theta,\varphi\right)=\frac{1}{2\pi^2\sin{\theta}}
\end{equation}
This equation was verified also by evaluating numerically Eq.\ (9) of reference \cite{catastrophes}, including summation over all branches. The function of Eq.\ (\ref{dist_analytic}) is plotted in Fig.\ \ref{z_kick_dist} by a dash-dotted line. Our numerical Monte Carlo result in Fig.\ \ref{z_kick_dist} (solid line) agrees well with the analytical result, but it is not singular at the north and the south poles, because a finite width $\sigma_{\text{belt}}$ was given to every molecule trajectory, and the singularities are smeared out.

Finally, we compare the classical result to the quantum angular distribution function averaged over a single revival period. The quantum result is given in Fig.\ \ref{z_kick_dist} by a dashed line. Two features should be stressed here. First, the singular classical behavior disappears in the quantum description. Second, there is a fine oscillatory structure in the quantum distribution, which is absent in the classical distribution. Besides these features, the distributions are similar. We also compare the values of $\overline{\langle x^2\rangle}$, $\overline{\langle y^2\rangle}$ and $\overline{\langle z^2\rangle}$, calculated both classically and quantum mechanically at the conditions shown in Fig.\ \ref{z_kick_dist} (b).  Here the brackets denote the ensemble averaging and the overline denotes long-time averaging. The quantities $x,y,z$ are the cartesian components of the unit vector $\mathbf{r}$. In both cases, the same values of
$\overline{\langle z^2\rangle}=0.42$ and $\overline{\langle x^2\rangle}=\overline{\langle y^2\rangle}=0.29$ are obtained. This should be compared to the "isotropic" values of $\overline{\langle x^2\rangle}=\overline{\langle y^2\rangle}=\overline{\langle z^2\rangle}=1/3$.

\begin{figure}
\centering
\includegraphics[width=1\textwidth]{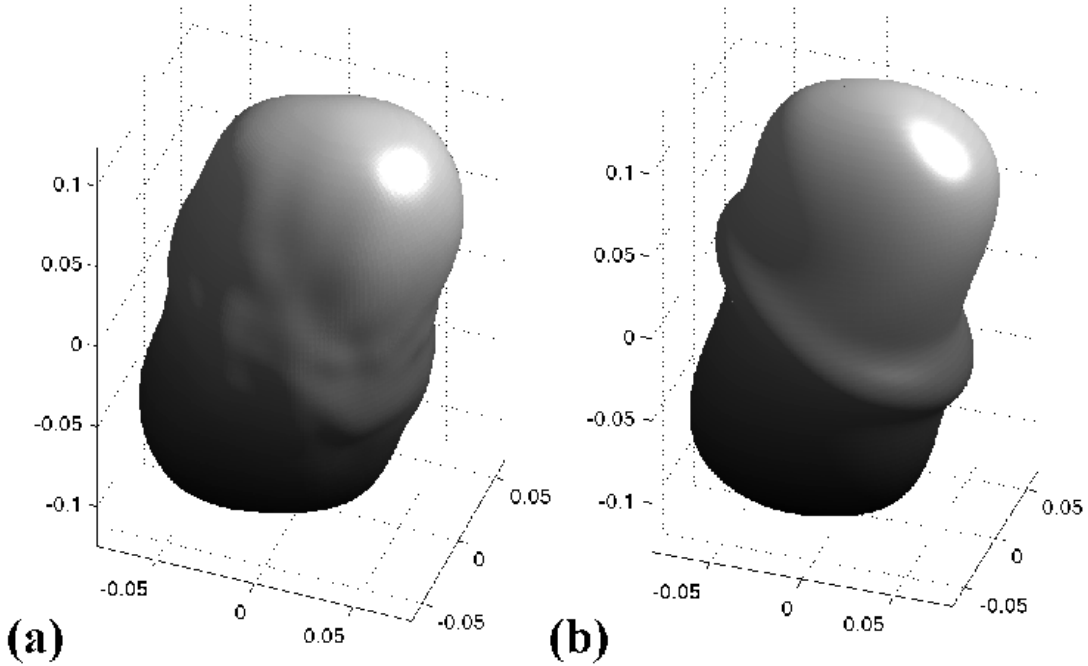}
\caption{Time-averaged angular distribution function for unidirectionally rotating nitrogen molecules that were excited by two delayed and cross-polarized pulses. The initial rotational temperature is $50\,\mathrm{K}$. The calculation was done using the classical Monte Carlo method, Eq.\ (\ref{time_avg_distribution}), in (a), and using the quantum finite-difference time-domain simulation followed by the  averaging over a revival period in (b). Both pulses had the kick-strength of $P=5$. The following parameters were used in (a): $\sigma_{\text{belt}}=0.1$, and $N=10^4$.}
\label{z_then45_dist}
\end{figure}

For the double pulse scheme, leading to the unidirectional rotation of the molecules, we present the classical time-averaged angular distribution in Fig.\ \ref{z_then45_dist} (a), which is very similar, except of fine details, to the corresponding quantum distribution in Fig.\ \ref{z_then45_dist} (b), presented also in \cite{njp}. The distribution is squeezed to the $xz$-plane (the plane defined by the polarization directions of the two pulses). We expected $\overline{\langle y^2\rangle}$ to be smaller than $\overline{\langle x^2\rangle}$ and $\overline{\langle z^2\rangle}$, and we indeed obtain numerically $\overline{\langle x^2\rangle}=0.31$, $\overline{\langle y^2\rangle}=0.30$ and $\overline{\langle z^2\rangle}=0.39$ classically, as well as quantum mechanically. These results demonstrate very good agreement between the classical and quantum calculations even for relatively small values of the interaction strength $P$.

\section{Unidirectional rotation of benzene molecules}

Now we address a more difficult problem of the unidirectional rotation of benzene molecules. These molecules, considered as rigid bodies, behave as oblate symmetric tops. The classical treatment of the free rotation of a symmetric top is analytically simple (although more complicated than the rotation of a linear rotor), and is well documented in literature (see, e.\ g.\ \cite[\S\S 33,35]{Landau}).

\subsection{Rotation of benzene molecules -- classical treatment}

\begin{figure}
\centering
\includegraphics[width=0.5\textwidth]{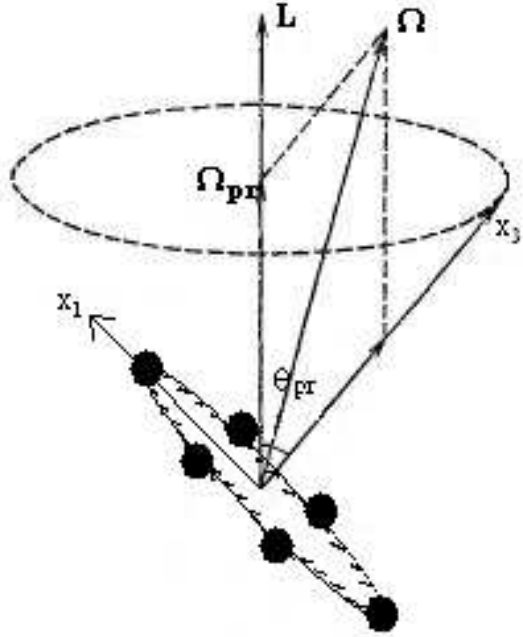}
\caption{The free rotation of a symmetric top with notation used in the text.}
\label{symmetric_top}
\end{figure}

There are two moments of inertia characterizing a symmetric top, $I_1=I_2$ and $I_3$, which are defined using the body fixed coordinate system $x_1,x_2,x_3$, shown in Fig.\ \ref{symmetric_top}, where the molecular symmetry axis is directed along the $x_3$ axis. The figure is taken from \cite[\S 33]{Landau}; the $x_2$ axis is perpendicular to the plane of the figure and is not shown. For benzene molecule, which is an oblate symmetric top, and has the form of a planar ring, the $x_3$ axis is perpendicular to the plane of the ring and passes through its center. Hence, for benzene, $I_1+I_2=2I_1=I_3$ \cite[\S 32]{Landau}.

When the molecule is freely rotating, the angular momentum, denoted by $\mathbf{L}$, is constant in magnitude and direction. The rotational kinetic energy is also a constant of motion and can be written as:
\begin{equation}
\label{Erot_sym_top}
T=\frac{L_1^2}{2I_1}+\frac{L_2^2}{2I_1}+\frac{L_3^2}{2I_3}=\frac{L_{\parallel}^2}{2I_1}+\frac{L_3^2}{2I_3}~,
\end{equation}
where $L_1$, $L_2$ and $L_3$ are the components of angular momentum along the body-fixed coordinate axes $x_1$, $x_2$ and $x_3$, respectively, and  $L_{\parallel}^2=L_1^2+L_2^2$.
The free rotation is such that the angular momentum $\mathbf{L}$, the angular velocity $\boldsymbol{\Omega}$ and the molecular axis $x_3$ all lie in the same plane. The molecular axis $x_3$ (as well as the vector $\boldsymbol{\Omega}$) performs precession around the angular momentum vector $\mathbf{L}$ with an angular velocity $\mathbf{\Omega_{pr}}$, which is shown in Fig.\ \ref{symmetric_top}. Its magnitude is:
\begin{equation}
\label{omega_pr}
\Omega_{\text{pr}}=\frac{L}{I_1}~.
\end{equation}
The molecular axis $x_3$ moves on the surface of a cone and forms an angle $\theta_{\text{pr}}$ with the cone axis (which is along the vector $\mathbf{L}$).

In addition to the precession motion, the top rotates around its $x_3$ axis, but this motion is not important in our problem. The reason is that this last rotation does not change the molecular orientation (defined by the molecular axis), and that the laser pulse does not influence this motion, as will be explained later on.

We describe the orientation of the molecular axis by a unit vector $\mathbf{r}$ in the direction of $x_3$. Its value at $t=0$ is $\mathbf{r_0}$. The velocity of this vector is given by
\begin{equation}
\label{v_benz_def}
\mathbf{v}=\mathbf{\Omega_{pr}\times\mathbf{r}}~.
\end{equation}
We now define two more unit vectors. One is $\vec{e}_{\mathbf{L}}$, a unit vector in the direction of $\mathbf{L}$. Another is in the plane of the circle traced by the tip of $\mathbf{r}$:
\begin{equation}
\label{r0_par}
\mathbf{r_{0\parallel}}=\frac{\mathbf{r_0}-\cos{\theta_{\text{pr}}}\vec{e}_{\mathbf{L}}}{\sin{\theta_{\text{pr}}}}~.
\end{equation}
Using the notation defined, it is easy to see that the time dependence of $\mathbf{r}$ is given by:
\begin{equation}
\label{r_benz}
\mathbf{r}(t)=\cos{\theta_{\text{pr}}}\vec{e}_{\mathbf{L}}+\sin{\theta_{\text{pr}}}\left(\mathbf{r_{0\parallel}} \cos{\Omega_{\text{pr}}t}+\frac{\mathbf{v_0}}{v_0}\sin{\Omega_{\text{pr}}t}\right)
\end{equation}
For the velocity we obtain:
\begin{equation}
\label{v_benz}
\mathbf{v}(t)=\Omega_{\text{pr}}\sin{\theta_{\text{pr}}}\left(-\mathbf{r_{0\parallel}} \sin{\Omega_{\text{pr}}t}+\frac{\mathbf{v_0}}{v_0}\cos{\Omega_{\text{pr}}t}\right)
\end{equation}
It is important to stress, that the tip of the unit vector $\mathbf{r}$ does not necessarily trace a  circle of a unit radius on the surface of the unit sphere, as for the linear molecule. It can trace any circle, which can even shrink to a point, in the case of $\Omega_1=\Omega_2=0$ and $\Omega_3\ne 0$. A unit circle is traced only if $\theta_{\text{pr}}=\pi/2$, or $\Omega_3=0$ and $\Omega_1=L/I_1$.

Although we are dealing here with a more complicated molecule, the interaction with an ultrashort laser pulse is described by the same formula as in Eq.\ (\ref{V}). The reason is that the polarizability tensor of the symmetric top has the same symmetry as the one of the linear molecule. The interaction depends, as before, on the electric field envelope squared, on $\Delta\alpha$, which is negative in the case of benzene, and on the angle between the pulse polarization direction $\mathbf{p}$ and the molecular axis orientation $\mathbf{r_0}$. Notice that a linearly polarized laser pulse does not produce a torque {\it parallel} to the molecular axis.

The torque exerted by the pulse is, therefore:
\begin{equation}
\label{torque}
\boldsymbol{\Gamma}=-\frac{1}{4}\mathcal{E}^2(t)\Delta\alpha\,\sin{2\beta_0}\,\vec{e}_{\mathbf{p\times r_0}}~,
\end{equation}
where $\vec{e}_{\mathbf{p\times r_0}}$ is a unit vector in the direction of the cross product $\mathbf{p\times r_0}$. This equation shows that the laser pulse ``kicks'' the symmetry axis of the benzene molecule towards the plane, which is perpendicular to the polarization vector $\mathbf{p}$. Using Newton's second law and integrating over the short time of the pulse, we obtain for the change in the angular momentum:
\begin{equation}
\label{dLx}
\Delta L_x=-\hbar P\sin{2\beta_0}\left(\vec{e}_{\mathbf{p\times r_0}}\right)_x~,
\end{equation}
and two similar equations for the $y$- and the $z$-components. Angle $\beta_0$ can be found using $\mathbf{p\cdot r_0}=\cos{\beta_0}$. Now, after the new angular momentum vector is known, and assuming the molecule did not move during the interaction with the pulse, the molecular rotation is completely defined by Eq.\ (\ref{r_benz}). Naturally, new values of $\theta_{\text{pr}}$, $\vec{e}_{\mathbf{L}}$, $\mathbf{r_{0\parallel}}$, $\mathbf{v_0}$ and $\Omega_{\text{pr}}$ should be found using the new value of $\mathbf{L}$. Specifically, $\theta_{\text{pr}}$ is found using the relation $\cos{\theta_{\text{pr}}}=\vec{e}_{\mathbf{L}}\cdot\mathbf{r_0}$, $\mathbf{r_{0\parallel}}$ is then defined  using Eq.\ (\ref{r0_par}), $\Omega_{\text{pr}}$ -- by Eq.\ (\ref{omega_pr}) and $\mathbf{v_0}$ -- using Eq.\ (\ref{v_benz_def}).

The Monte Carlo simulation for benzene molecules has several  differences compared with the one used for linear molecules. Here, the kinetic energy is more easily separable when expressed using the angular momentum components along the body-fixed coordinate system, as in Eq.\ (\ref{Erot_sym_top}).

We start with random generation of the angular momentum components $L_{\parallel}$ and $L_3$, according to the thermal distribution:
\begin{eqnarray}
\label{boltzmann_benzene}
\text{Prob}\left(L'_{\parallel}\right) dL'_{\parallel} &\times& \text{Prob}\left(L'_3\right) dL'_3 \\
~&=& \frac{1}{\sigma_{\text{th},1}^2} \exp{\left[-\frac{\left(L'_{\parallel}\right)^2}{2\sigma_{\text{th},1}^2}\right]} L'_{\parallel}\, dL'_{\parallel} \nonumber \\
~&\times & \frac{1}{\sqrt{2\pi\sigma_{\text{th},3}^2}} \exp{\left[-\frac{\left(L'_3\right)^2}{2\sigma_{\text{th},3}^2}\right]} dL'_3~,\nonumber
\end{eqnarray}
where we introduce $\sigma_{\text{th},1}^2=I_1k_BT/\hbar^2$ and $\sigma_{\text{th},3}^2=I_3k_BT/\hbar^2$, similar to Eq.\ (\ref{boltzmann}), and where $L'$ represents the dimensionless angular momentum $L/\hbar$. Employing the transformation method, and using the notation introduced after Eq.\ (\ref{rand_r_sphere}), we obtain explicitly for the random components: $(L'_{\parallel})^{\text{rand}}=\sqrt{2}\,\sigma_{\text{th},1} \sqrt{\ln{[1/(1-w_{\text{un}})]}}$ and $(L'_3)^{\text{rand}}=\sigma_{\text{th},3} w_{\text{norm}}$. Knowing these components, we deduce the {\it magnitude} of the angular momentum $L=\sqrt{L_{\parallel}^2+L_3^2}$, but  its {\it direction} is yet to be defined. We obtain also the precession velocity using Eq.\ (\ref{omega_pr}), and $\theta_{\text{pr}}$ from the relation $\cos{\theta_{\text{pr}}}=L_3/L$.

Next, we determine the {\it direction} of the angular momentum in the laboratory coordinate system, which is given by the vector $\vec{e}_{\mathbf{L}}$. We do this by randomly generating two spherical angles $\theta_\mathbf{L}$ and $\varphi_\mathbf{L}$. These two angles are distributed the same as $\theta$ and $\varphi$ in Eq.\ (\ref{rand_r_sphere}), because the tip of the vector $\vec{e}_{\mathbf{L}}$ is distributed uniformly on the surface of the unit sphere.

Finally, by randomly generating an angle $\varphi$, uniformly distributed between $0$ and $2\pi$, we obtain the initial orientation of the molecular axis on the cone around the vector $\mathbf{L}$. Now, the coordinates of $\mathbf{r_0}$ are completely defined in the frame in which the $z$-axis is  fixed along the vector $\mathbf{L}$. These coordinates are given by $(\sin{\theta_{\text{pr}}}\cos{\varphi}, \sin{\theta_{\text{pr}}}\sin{\varphi}, \cos{\theta_{\text{pr}}})$. Using two rotational matrices, the coordinates of $\mathbf{r_0}$ in the laboratory frame are then found.

The above Monte Carlo procedure was used to generate data that are discussed later in Sec. C.

\subsection{Quantum mechanical treatment}

Here we outline the quantum mechanical description of unidirectional rotation of benzene molecules by two delayed ultrashort laser pulses with  crossed linear polarizations.  Details will be presented in \cite{Kitano2010}.  The molecular Hamiltonian, $ \hat{H}_{0} $, in the field-free conditions is obtained simply by replacing the components, $ L_{i}$ $(i = 1,2,3) $, of the classical angular momentum by the corresponding quantum mechanical operators, $ \hat{J}_{i} $, in Eq.\ (\ref{Erot_sym_top}).  The rotational eigenstates and eigenenergies are given by the time-independent Schr$ \rm {\ddot{o}}$dinger equation,
\begin{equation}
\label{TISE}
\hat{H}_{0} \vert r \rangle = E_{r} \vert r \rangle ~,
\end{equation}
where $r$ stands for an index to identify the eigenstates, explicitly expressed by a set $\{J, K, M\}$, with $J$ being the total angular momentum, and $K$ and $M$ being its projection onto the molecular symmetry axis and the space-fixed axis, respectively.
At finite temperature, an initial ensemble is a mixture of molecules in many different eigenstates, but we first take a single rotational level as an initial state, $ \vert r_{i} \rangle  $. The nonadiabatic interaction with the nonresonant ultrafast laser field, given by  Eq.\ (\ref{V}), converts the stationary state to a rotational wave packet, $ \vert \Psi (t) \rangle  $, which is expanded as
\begin{eqnarray}
\label{Psi}
\vert \Psi_{r_{i}} (t) \rangle = \hat{U}(t,0) \vert r_{i} \rangle = \sum_{r} C_{r_{i},r}(t) \exp (-i \omega_{r} t) \vert r \rangle ~.
\end{eqnarray}
Here, $ \hat{U} (t_{2}, t_{1}) $ is the time-evolution operator from time $ t_{1} $ to $ t_{2} $ and $ \omega_{r} = {E_{r} / \hbar} $.  The expansion coefficients, $C_{r_{i},r}(t)$, vary during the interaction with the laser pulses, and subsequently become constant after the laser field vanishes.  They can be derived by solving the time-dependent Schr$ \rm {\ddot{o}}$dinger equation (TDSE) with the initial condition, $ \vert \Psi (t \rightarrow - \infty) \rangle = \vert r_{i} \rangle $,
\begin{eqnarray}
\label{TDSE}
i \hbar \frac{\partial}{\partial t} \vert \Psi_{r_{i}} (t) \rangle = \left[ \hat{H}_{0} + \hat{V} (t) \right]  \vert \Psi_{r_{i}} (t) \rangle ~,
\end{eqnarray}
where $ \hat{V} (t) $ is given in Eq.\ (\ref{V}).  Substituting Eq.\ (\ref{Psi}), this TDSE is recast into the following coupled differential equations for the expansion coefficients:
\begin{eqnarray}
\label{CDE}
i\hbar \frac{d}{dt} C_{r_{i},r} (t) = \sum_{r'} \langle r \vert \hat{V} (t) \vert r' \rangle \exp (-i \Delta \omega_{r',r} t)  C_{r_{i},r'} (t) ~,
\end{eqnarray}
with $ \Delta \omega_{r',r} =  \omega_{r'} - \omega_{r}$.  These coupled equations can be solved numerically to determine the complex expansion coefficients, once the matrix elements for $ \hat{V}$ are specified.

If another laser pulse is irradiated onto the molecule at $t = \tau$, the wave packet created by the first pulse, represented in Eq.\ (\ref{Psi}), is further modified by the interaction with the second pulse.  The resultant wave packet at $t$ after the second pulse is expanded as
\begin{eqnarray}
\label{Psi_new}
\vert \Psi_{r_{i}} (t) \rangle = \hat{U'}(t,\tau) \vert \Psi_{r_{i}} (\tau) \rangle = \sum_{r} B_{r_{i},r} (\tau) \exp (-i \omega_{r} t) \vert r \rangle ~,
\end{eqnarray}
where the prime on $\hat{U}$ indicates the properties of the second pulse. The quantity
$ B_{r_{i},r} $ is the transition amplitude from the initial $ \vert r_{i} \rangle $ state to $ \vert r \rangle $ by the interaction with the two pulses, and is represented by
\begin{eqnarray}
\label{B}
B_{r_{i},r} (\tau) = \sum_{r'} C_{r_{i},r'} C'_{r',r} \exp (-i \Delta\omega_{r',r} \tau) ~.
\end{eqnarray}

To evaluate the transition amplitude given in Eq.\ (\ref{B}), it is convenient to adopt the space-fixed axis system defined differently from the previous one.  Here, we set the $z$ axis along the laser propagation direction and the polarization of the first pulse parallel to the $x$ axis.  Then, the interaction given in Eq.\ (\ref{V}) is recast as \cite{Ohshima2010}
\begin{eqnarray}
\hat{V} ( \theta,\varphi, t) &=& \frac{1}{8}\mathcal{E}^2(t) \left\{ \Delta\alpha \left[1+\cos(2 \varphi) \right] \cos^2{\theta} \right. \nonumber \\
\label{V_new}
~&-&\left. \Delta\alpha \cos(2 \varphi) - ( \alpha_{\parallel} + \alpha_{\perp} ) \right\} ~,
\end{eqnarray}
or, re-expressed as
\begin{eqnarray}
\hat{V} &=&-\frac{1}{12}\mathcal{E}^2(t) \left\{ \left(\alpha_{\parallel } + 2\alpha_{\perp} \right) - \Delta\alpha D^{(2)*}_{0,0} \right.\nonumber \\
\label{V_new2}
~&+&\left. \sqrt{\frac{3}{2}} \Delta\alpha \left[D^{(2)*}_{-2,0} + D^{(2)*}_{2,0} \right]  \right\} ~,
\end{eqnarray}
where $D^{(j)}_{p,q} $ is the rotational matrix \cite{Zare}.  With this expression, the evaluation of the matrix elements for a symmetric-top basis set is straightforward by applying the spherical tensor  algebra, and the following selection rules are derived:

$\Delta K = 0$,

$\Delta M = 0$ (then $\Delta J = 0, \pm2$ for $KM = 0$ and $\Delta J = 0, \pm1, \pm2$ for $KM\neq0$), or

$\Delta M = \pm2$ (then $\Delta J = 0, \pm2$ for $K = 0$ and $\Delta J = 0, \pm1, \pm2$ for $K\neq0$).

We set the polarization of the second pulse tilted against that of the first one by angle $\Delta \varphi $.  When the new axis system is defined so as the $x'$ axis is set parallel to the polarization of the second pulse, the new and old angular coordinates are related as: $ \varphi' = \varphi - \Delta \varphi $.  Consequently, the symmetric-top wave function transforms as
\begin{equation}
\label{transform}
\vert J,K,M \rangle \longmapsto \vert J,K,M \rangle \exp (i M \Delta \varphi )~.
\end{equation}
By taking the transformation into account, $ B_{r_{i},r} $ in Eq.\ (\ref{B}) is rewritten, if the second pulse is a replica of the first one, as
\begin{eqnarray}
\label{B_new}
B_{r_{i},r} (\tau) = \sum_{r'} C_{r_{i},r'} C_{r',r} \exp \left[ (-i \Delta\omega_{r',r} \tau - M' \Delta \varphi ) \right] ~,
\end{eqnarray}
with $r'$ standing for a set of $\{J', K', M'\}$.
The transition amplitude for the two pulses can be evaluated for an arbitrary $\tau$ at once, if the set of the coefficients, $C_{r_{i},r'}$, are derived by solving the coupled equations in (\ref{CDE}).

The expectation value of a given observable $\hat{A}$ for a certain initial condition with finite temperature is derived as the following ensemble average,
\begin{eqnarray}
\label{average}
\langle \hat{A} \rangle (\tau,t) &=& \sum_{r_{i}} W_{r_{i}} \langle \Psi_{r_{i}} (t) \vert \hat{A} \vert \Psi_{r_{i}} (t) \rangle \\
~&=& \sum_{r_{i},r,r'} W_{r_{i}} B^{*}_{r_{i},r'} (\tau) B_{r_{i},r} (\tau) \langle r' \vert \hat{A} \vert r \rangle \exp (i \Delta\omega_{r',r} t) ~,\nonumber
\end{eqnarray}
where $W_{r_{i}}$ is the Boltzmann factor for $\vert r_{i} \rangle$ with appropriate degeneracy factor due to nuclear-spin statistics \cite{Beck}.
For the discussion of the molecular alignment and orientation, we take $\cos^2 \theta$, $\hat{J}^2$, and $\hat{J}_z$ (corresponding to the classical $L_y$), as $\hat{A}$.

\subsection{Classical vs quantum treatment}

In this Section we analyze the process of exciting unidirectional rotation of benzene, following the setup of \cite{Kitano}. The scheme operation is practically the same as described after Eq.\ (\ref{P}), except of several details specific for the benzene molecules.
Shortly after the first ultrashort pulse, polarized along $\mathbf{p_1}=(0,0,1)$, the molecules experience a transient anti-alignment, as opposed to alignment in the case of linear molecules. This happens because $\Delta\alpha$ is negative for benzene, as opposed to linear molecules, such as nitrogen, and the direction of the angular momentum transferred to the molecule by the pulse is reversed. Therefore, shortly after the first pulse the molecular axis angular distribution is squeezed to the plane {\it perpendicular} to $\mathbf{p_1}$. Quantitatively, after the first pulse, the ensemble average of $\langle\cos^2{\theta}\rangle$ decreases below the isotropic value of $1/3$. The second delayed and tilted laser pulse brings this anisotropic ensemble to the rotation with a preferred sense. In the following, we shall consider the time development of the alignment factor $\langle\cos^2{\theta}\rangle$ after the first pulse, and the dependence of the average value of the induced angular momentum as the function of the delay of the second pulse.

\begin{figure}
\centering
\includegraphics[width=1\textwidth]{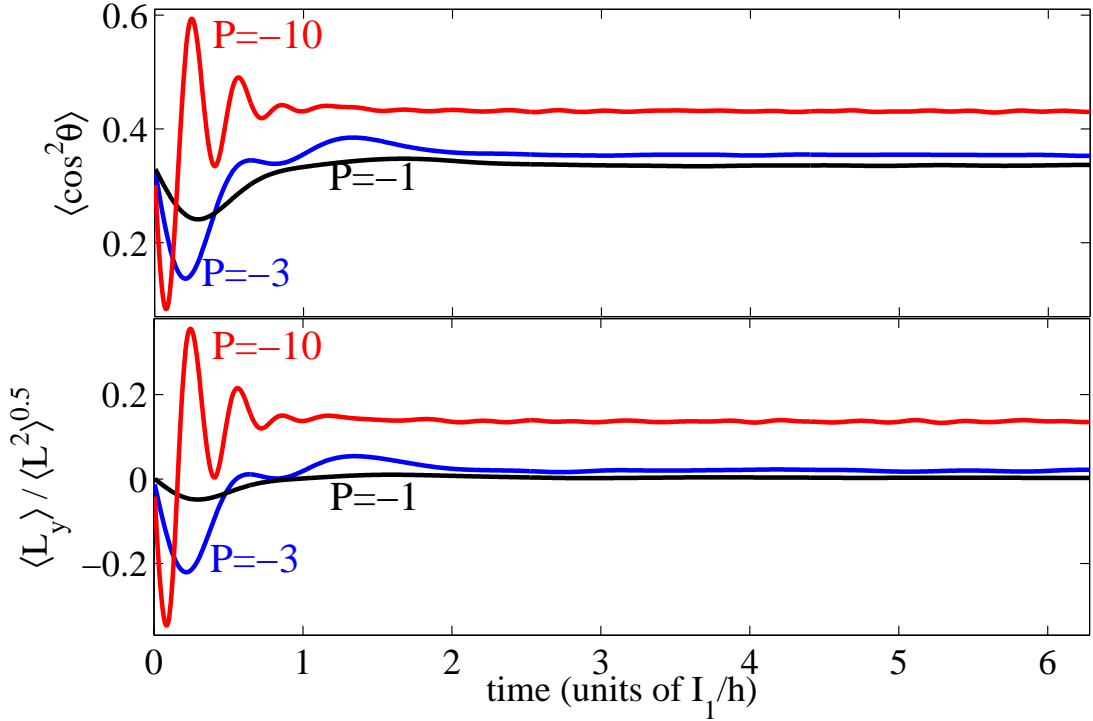}
\caption{(Color online) Classical treatment. The upper panel shows thermally averaged alignment factor $\langle\cos^2{\theta}\rangle$ for benzene molecules kicked by a single pulse,  plotted as a function of time after the pulse. The pulse has a strength of $P_1=-1,-3,-10$ (black, blue and red lines online, respectively) and  polarization direction  $\mathbf{p_1}=(0,0,1)$. The lower panel displays the (normalized) value of the oriented angular momentum  $\langle L_y\rangle/\sqrt{\langle L^2\rangle}$ induced by a second ultrashort pulse, shown as a function of the delay of this pulse. The polarization direction of the second pulse is $\mathbf{p_2}=(-1,0,1)/\sqrt{2}$ (at $45$ degrees to $\mathbf{p_1}$), and the value of $P_2$ is the same as that of $P_1$, with the same color code used for the curves. The initial rotational temperature is $0.9\,\text{K}$. The calculations were done using the Monte Carlo method with $10^5$ realizations. }
\label{benzene_Jy_cos2_class}
\end{figure}

We show first the classical results. The upper part of Fig.\ \ref{benzene_Jy_cos2_class} presents the alignment factor $\langle\cos^2{\theta}\rangle$ as a function of time passed after the first pulse (polarized along $\mathbf{p_1}=(0,0,1)$), for three values of the pulse  strength $P$. Here and below we use the following parameters for the benzene molecule \cite{Hasegawa}: rotational constants $B=h/(8\pi^2I_1c)=0.190\,\mathrm{cm}^{-1}$, $C=h/(8\pi^2I_3c)=B/2$  and the polarizabilities $\alpha_{\parallel}=6.67\,\text{\r{A}}^3$, $\alpha_{\perp}=12.4\,\text{\r{A}}^3$. All plots  are for rotational temperature of $0.9\,\text{K}$ (as in experiment \cite{Kitano}),  which corresponds to the dimensionless parameters $\sigma_{\text{th},1}=1.29$ and $\sigma_{\text{th},3}=\sqrt{2}\sigma_{\text{th},1}=1.82$.

\begin{figure}
\centering
\includegraphics[width=1\textwidth]{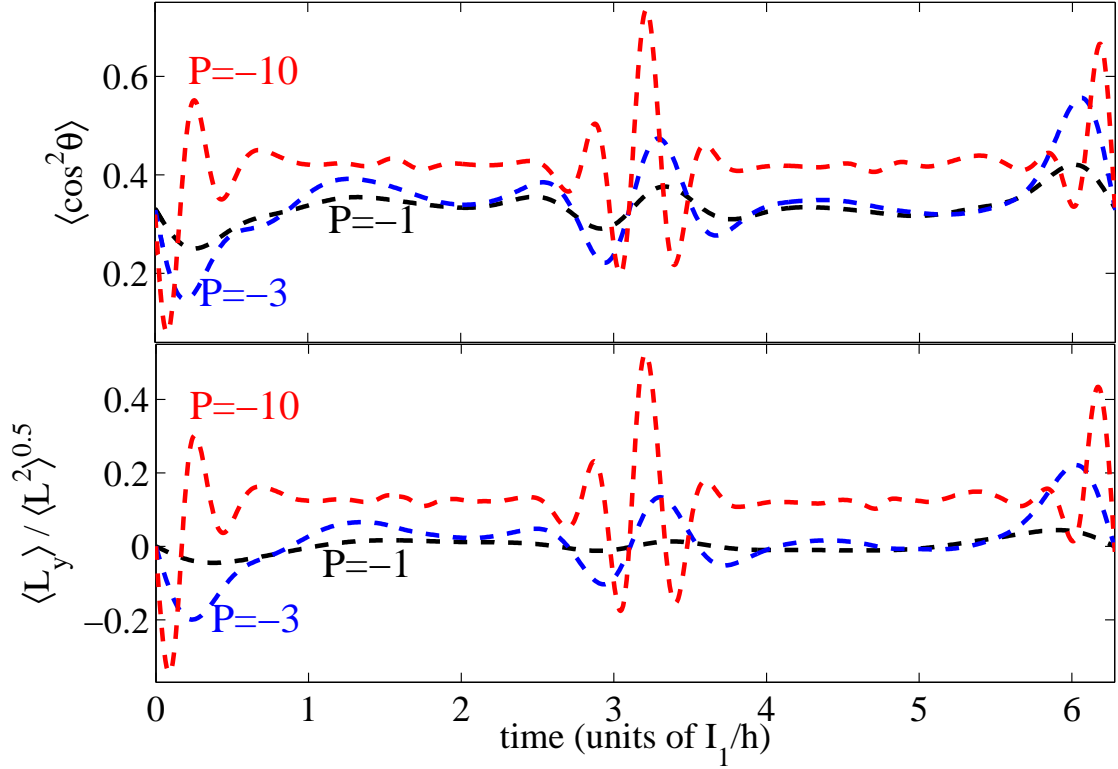}
\caption{(Color online) The same quantities as in Fig.\ \ref{benzene_Jy_cos2_class} are plotted here. They were calculated using the quantum mechanical treatment. Half and full rotational revivals are seen, which are not present in Fig.\ \ref{benzene_Jy_cos2_class}.}
\label{benzene_Jy_cos2_quant}
\end{figure}

\begin{figure}
\centering
\includegraphics[width=1\textwidth]{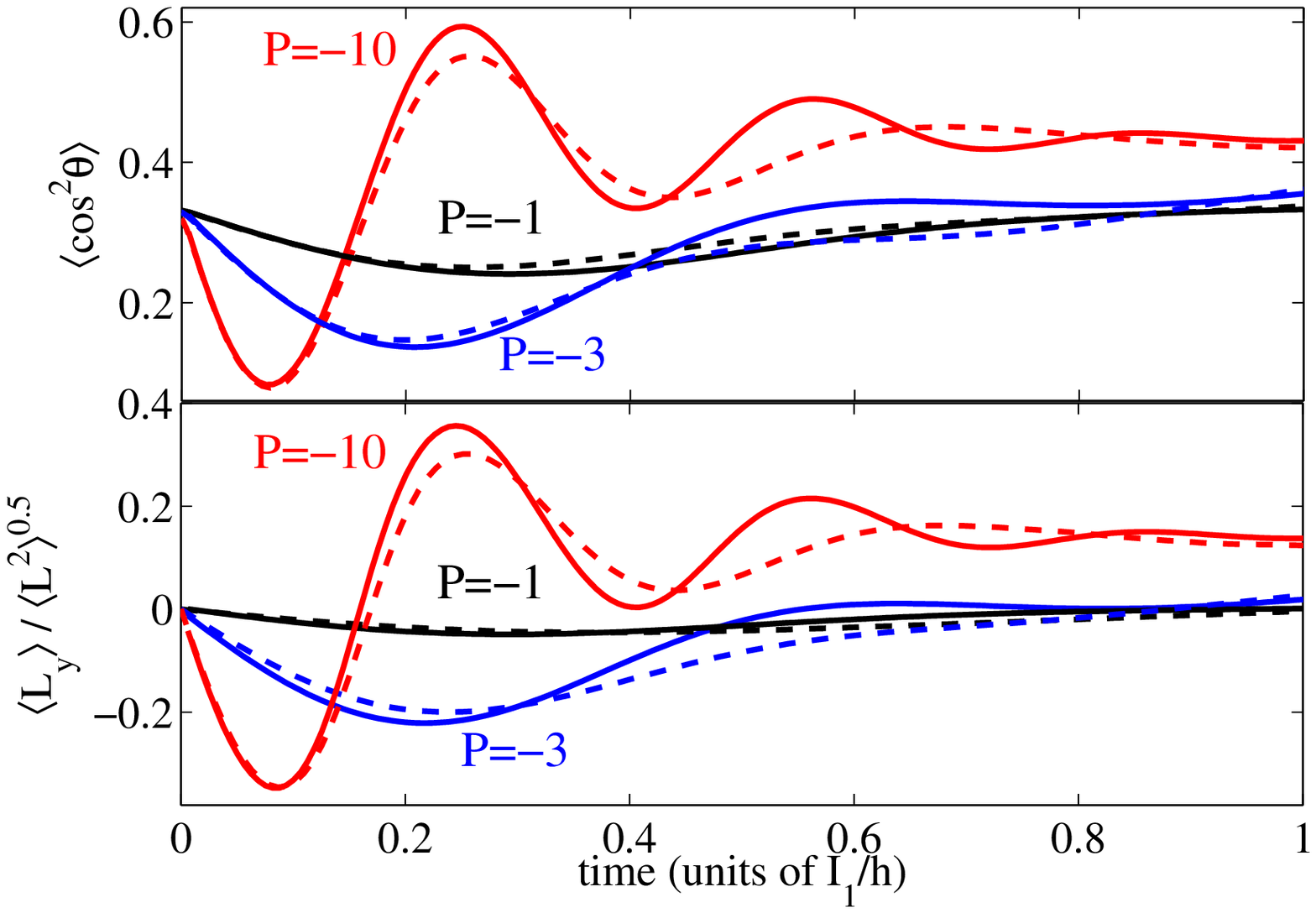}
\caption{(Color online) Comparison of the classical results of Fig.\ \ref{benzene_Jy_cos2_class} (solid lines), and  quantum mechanical results of Fig.\ \ref{benzene_Jy_cos2_quant} (dashed lines) at short times. The values of $P$ are $-1,-3,-10$ for black, blue and red lines, respectively.  }
\label{benzene_Jy_cos2_compar}
\end{figure}

The plots for $\langle\cos^2{\theta}\rangle$ show that the anti-alignment becomes more pronounced and occurs at earlier times as the pulse intensity increases,  as is naturally expected. Similar to the case of  linear diatomic molecules, we anticipate that the second pulse should be fired at the  moment of the maximal squeezing of the molecular angular distribution, i.\ e.\ when the alignment factor takes the minimal value. In this case, the  maximal average torque is applied to the molecules, facilitating the most emphasized unidirectional rotation. This is indeed seen from the lower part of Fig.\ \ref{benzene_Jy_cos2_class}. Here the quantity $\langle L_y\rangle/\sqrt{\langle L^2\rangle}$ (calculated after the second pulse) is plotted as a function of the time delay between the first pulse (with polarization direction $\mathbf{p_1}$) and the second pulse with polarization direction $\mathbf{p_2}=(-1,0,1)/\sqrt{2}$, where the angle between $\mathbf{p_1}$ and $\mathbf{p_2}$ is $45$ degrees. The maximal modulus of the transferred angular momentum  happens indeed when the second pulse is applied at the moment of the maximal anti-alignment. We also notice that the optimal delay becomes shorter with the strength of the pulses.

Figure \ref{benzene_Jy_cos2_quant} presents the same quantities calculated fully quantum mechanically, following the procedure outlined in section V-B. On the long-time scale, it demonstrates quantum rotational revivals, such as  half and  full revivals, that are absent in Fig.\  \ref{benzene_Jy_cos2_class} for obvious reasons. For a detailed comparison with the classical results at shorter times (corresponding to the conditions of the experiment \cite{Kitano}),  Fig.\  \ref{benzene_Jy_cos2_compar} displays the classical and the quantum graphs overlapped on the same plot. The results seem to be practically identical at short times, and the agreement is improving with the increase of $|P|$, as more rotational states are involved in the dynamics of the wavepacket. Parameters of the experiment \cite{Kitano} correspond to the curves for $P=-3$ in Fig.\  \ref{benzene_Jy_cos2_compar}, with the delay time chosen near the minima of the curves. As follows from Fig.\  \ref{benzene_Jy_cos2_compar}, the rotational dynamics in this range  is adequately described by the classical treatment.

\section{Conclusions}

We studied the problem of exciting unidirectional  molecular rotation with the help of two delayed cross-polarized laser pulses using classical mechanics, and compared this approach with the results of fully quantum mechanical treatment. The problem was analyzed both for linear molecules and for symmetric tops, like benzene.  Although the comprehensive description of the fine details of rotational dynamics undeniably requires the full scale quantum formulation, we found a very good agreement between the classical and quantum approaches in a wide range of experimentally relevant parameters. The classical model identifies the origin of the mechanism, defines the optimal delay between the pulses, and the best angle between them. Moreover, it accurately predicts the magnitude of the induced oriented angular momentum, and correctly reproduces the behavior of the optimal delay with pulse strength, as it was observed in \cite{Kitano}. At the same time, a fully quantum treatment is required to describe quantitatively the operation of the double-pulse scheme for relatively weak laser pulses, when only low rotational states are involved. In this case, considering quantum interference of excitation pathways by the first and the second pulses \cite{Kitano} provides an instructive complementary view on the origin of the phenomenon. Moreover, operation of the scheme at long delays between the pulses (comparable with the period of rotational revivals) needs quantum description as well. At the same time, quantum and classical results agree well again if the long-time-averaged observable quantities are considered. We illustrated this by comparing classical and quantum time-averaged molecular angular distributions that show anisotropy and confinement to the plane spanned by two polarization vectors of the laser pulses.

We believe that this comparative study provides the reader a comprehensive view on the physics behind the double pulse scheme for ultrafast preparation of molecular ensembles with oriented rotational angular momentum and controlled sense of rotation.

\section{Acknowledgment}

The authors are thankful to Kenji Ohmori for stimulating this collaborative work. YK and IA appreciate many fruitful discussions with Yehiam Prior, Sharly Fleischer and Erez Gershnabel. Financial support of this research by the Israel Science Foundation is gratefully acknowledged.
Additional financial supports by a Grant-in-Aid from MEXT Japan, C-PhoST, and ``Extreme Photonics" are also appreciated.
IA is an incumbent of the Patricia Elman Bildner Professorial Chair. This research is made possible in part by the historic generosity of the Harold Perlman Family.

\end{document}